# Rapid Prototyping of 3D Microstructures: A Simplified Grayscale Lithography Encoding Method Using Blender

**Fabrício Frizera Borghi[1]\***, **Mohammed Bendimerad[2,3]**, **Marie-Ly Chapon[2,3]**, **Tatiana Petithory[2,3]**, **Laurent Vonna[2,3]**, and **Laurent Pieuchot[2,3]\***

1: Instituto de Fisica, Universidade Federal do Rio de Janeiro, Rio de Janeiro 21941-611, Brasil

2: Université de Haute-Alsace, CNRS, IS2M UMR 7361, Mulhouse 68100, France

3: Université de Strasbourg, Strasbourg, F-67081, France

\*For correspondance: borghi@if.ufrj.br; laurent.pieuchot@uha.fr

**Abstract**

The democratization of fabrication equipment has spurred recent interest in maskless grayscale lithography for both 2D and 3D microfabrication. However, the design of suitable template images remains a challenge. This work presents a simplified method for encoding 3D objects into grayscale image files optimized for grayscale lithography. Leveraging the widely used, open-source 3D modeling software Blender, we developed a robust approach to convert geometric heights into grayscale levels and generate image files through top-view rendering. Our method accurately reproduced the overall shape of simple structures like stairs and ramps compared to the original designs. We extended this approach to complex 3D sinusoidal surfaces, achieving similar results. Given the increasing accessibility and user-friendliness of digital rendering tools, this study offers a promising strategy for rapid prototyping of initial designs with minimal effort.

## 1. Introduction

Grayscale lithography is a technique that enables the creation of three-dimensional microstructures in light-sensitive photoresist. It involves exposing a photoresist to varying light intensities, resulting in different depths after development (Figure 1a). This method has been employed to fabricate microelectromechanical systems (MEMS), microlens arrays, Fresnel lenses, and molds for soft-lithography (Figure 1c) [1,2,3].

Traditionally, this was achieved using complex sets of physical hard masks, which were sequentially applied to expose specific regions to varying light levels. However, this approach suffered from poor control, limited spatial resolution, high costs, lengthy fabrication times, and inflexibility, hindering its widespread adoption [4-7].

To overcome these limitations, maskless lithography has emerged as a viable alternative. This technology utilizes digital designs to precisely control light exposure based on desired dose and region. There are two primary maskless lithography methods: Direct Write Lithography (DWL), which employs a focused, power-controlled laser and precise stage positioning [8]; and Digital Mirror Device (DMD) technology, which uses microscopic mirrors to modulate light intensity and distribution [9] (Figure 1b).

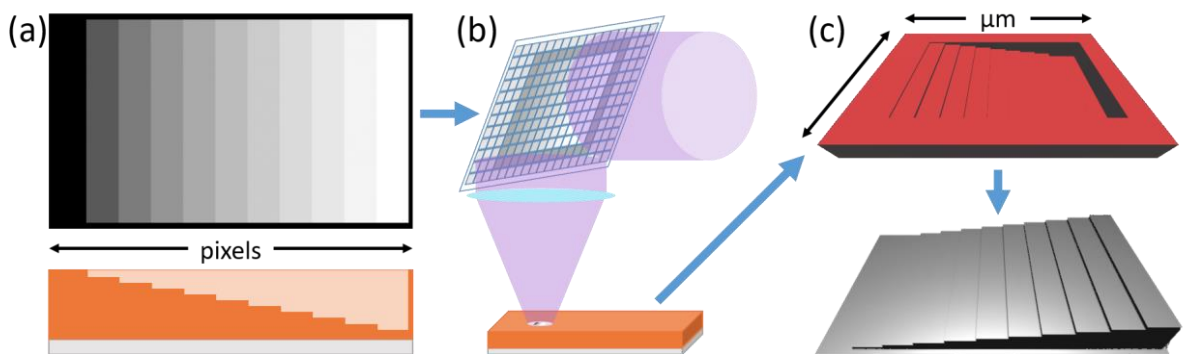

**Figure 1.** (a) Digital image in grayscale and corresponding material removal after development. Brighter gray levels correspond to higher applied power, resulting in greater material removal. (b) The image-based instructions are converted into mirror positioning on the DMD and focused onto the photoresist with appropriate intensity. (c) Representations of the expected results for the mold and soft-lithography replication.

Although termed maskless due to the absence of a physical mask between the light source and photoresist, direct laser writing (DWL) necessitates digital instructions (image or 3D files) to accurately position the substrate and control laser power. In contrast, digital micromirror device (DMD) technology replaces the physical mask by employing an array of micromirrors to modulate both light intensity and position on the photoresist surface. Pioneered by Takahashi and Setoyama in 1999, DMD-based UV exposure systems achieved initial line/space resolutions of 50 µm [10]. Beyond binary patterns, DMD enables the creation of grayscale masks through computer control, finding applications in printed circuit boards (PCBs), microelectromechanical systems (MEMS), microlenses, and 2D/3D microstructures [11, 12]. Similar to DWL, DMD processes rely on digital instructions, typically grayscale images, to regulate light transmission and exposure on the photoresist.

Several methods have been proposed to generate grayscale lithography masks for three-dimensional microstructures. The first approach consists in using specific software, either open source or proprietary, to slice and transform 3D shapes into images. Most of the reported shapes are simple, created using a script, and prepared in an image format [13]. Loomis et. al. [1] described an automated process and the corresponding grayscale conversion software that they developed to generate image files.

After generating the instruction files, Smith et al. [14] and Erjawetz et. al. [13] demonstrated the applicability of software to accurately simulate how designs will develop on photoresists by correcting for proximity effects during exposure and predicting the development. Both approaches rely on the a priori characterization of the photoresist used, and although they mention the generation of the digital mask, they mostly contribute to its redesign and microstructures shape correction.

Regardless of the technology used, they must account for the nonlinear relationship between light exposure and photoresist removal [4]. This can be done after calibration, by adjusting the gray levels at the machine using a look-up table (LUT) that converts the gray levels into the measured calibration curve. Alternatively, the correction can be performed on the original digital mask by adjusting the shapes of the desired microstructures and generating the new image file via dedicated conversion software or code.

This study aims to streamline the generation of image files for maskless grayscale lithography. We propose to use Blender, an open-source 3D modeling and rendering software, to design structures and generate corresponding height maps. An automated workflow, including exposure dosage calibration and camera view rendering for export, is presented. Complex grayscale height maps were designed and the resulting structures characterized using SEM, confocal microscopy, and profilometry. A comparative analysis between the fabricated structures and original models is provided.

## 2. Materials and Methods

### 2.1. Lithography

Standard microscope glass slides were cut into 2x2 cm² pieces and subjected to a cleaning process involving sequential ultrasonic baths in soapy distilled water, acetone, and isopropanol (2 minutes each), followed by air drying. A uniform 30 µm thick layer of ma-P 1275G photoresist (Micro Resist Technology) was deposited on the cleaned substrates through spin coating at 450 rpm for 60 seconds. Subsequent heating at 100 °C for 10 minutes on a hotplate ensured photoresist hardening and prevented bubble formation. The coated substrates were allowed to rehydrate at room temperature for at least 2 hours to facilitate the transition from ketene intermediate to carboxylic acid (ICA). Exposure was performed using a Smart Print UV (Microlight 3D, DMD projection) lithography system equipped with a x10 lens. Each digital mask was exposed for 5 seconds at 20% UV LED power (approximately 1920 mW/cm² at 385 nm). The exposed photoresist was developed in mr-D 526/S developer (Micro Resist Technology) for 3 minutes to dissolve the ICA and reveal the underlying structure. No post-baking was performed.

## 2.2. Calibration

A nonlinear relationship exists between gray level and photoresist removal, which was determined through individual calibration of 50 data points. Each calibration sample consisted of a 100 µm square filled with a single gray level (1-150 on a 255-level scale) selected to induce material removal exceeding 20 µm, our region of interest. Post-lithography profilometry measurements of these samples yielded the average removed material depth within each square. Plotting these depths against corresponding gray levels produced the contrast curve (Figure 2a). To map height to color in the 3D design, the contrast curve was inverted, normalized, and fitted (Figure 2b). This fitted curve was then applied as a color ramp within the rendering software, correlating gray levels to Z-axis depth (Figure 2c). Importantly, this calibration process can be adapted for different regions of interest and readily updated as needed.

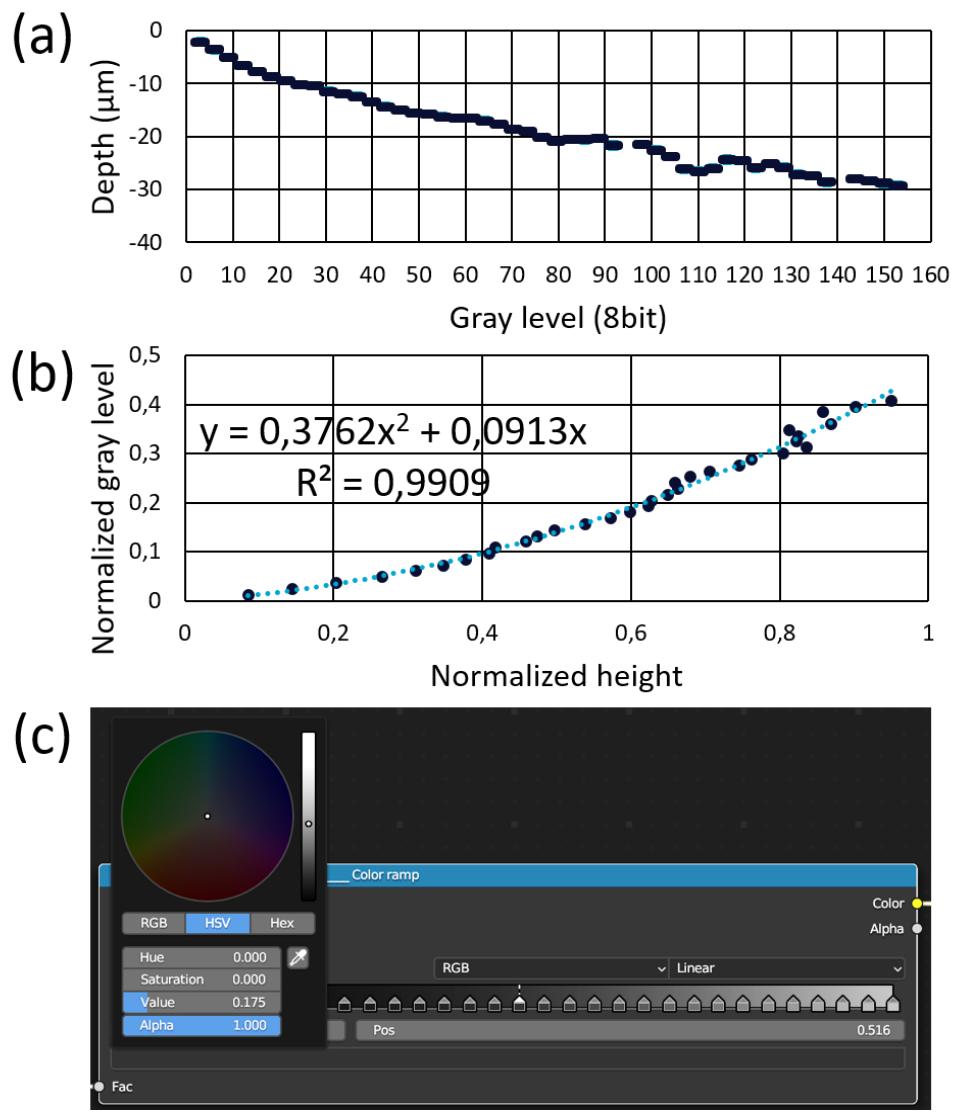

**Figure 2.** (a) Photoresist contrast curve determined by profilometry measurements following individual light power exposures corresponding to different gray levels. (b) Normalized and inverted contrast curve with fitted data. (c) Blender color ramp configured based on the calibration curve.

## 2.3. Designing the 3D microstructures

The open-source software Blender [15] was employed to generate digital mask image files. Renowned for its versatility in modeling, animation, rendering, simulation, video editing, and interactive 3D applications, Blender is widely utilized across design, gaming, and scientific visualization [16]. This software facilitated 3D design creation, height-based colorization, and perspective-free (Z-axis) rendering. To adapt Blender's capabilities for this application, several settings were adjusted. Color filters were reset for accurate display and rendering, image dimensions set to 1920x1080 pixels, and output format configured as grayscale TIFF (8-bit, uncompressed). A planar working area with dimensions converted to micrometers (960x540 µm²) was established. The camera was positioned above the plane in orthographic mode, ensuring a top-down view of the entire working area (Figure 3a-c). Height-based shading was implemented using a calibration curve to map Z-axis values to color gradients (Figure 3g). The resulting rendered images (Figure 3d-f) served as calibrated digital masks for arbitrary 3D designs. To illustrate this process, we created simple (ramp, stairs), complex (sinusoidal patterns), and geometric (pyramids, cones, cylinders, cubes, hemispheres) structures, each measuring 250x250 µm² with a maximum depth of 15 µm.

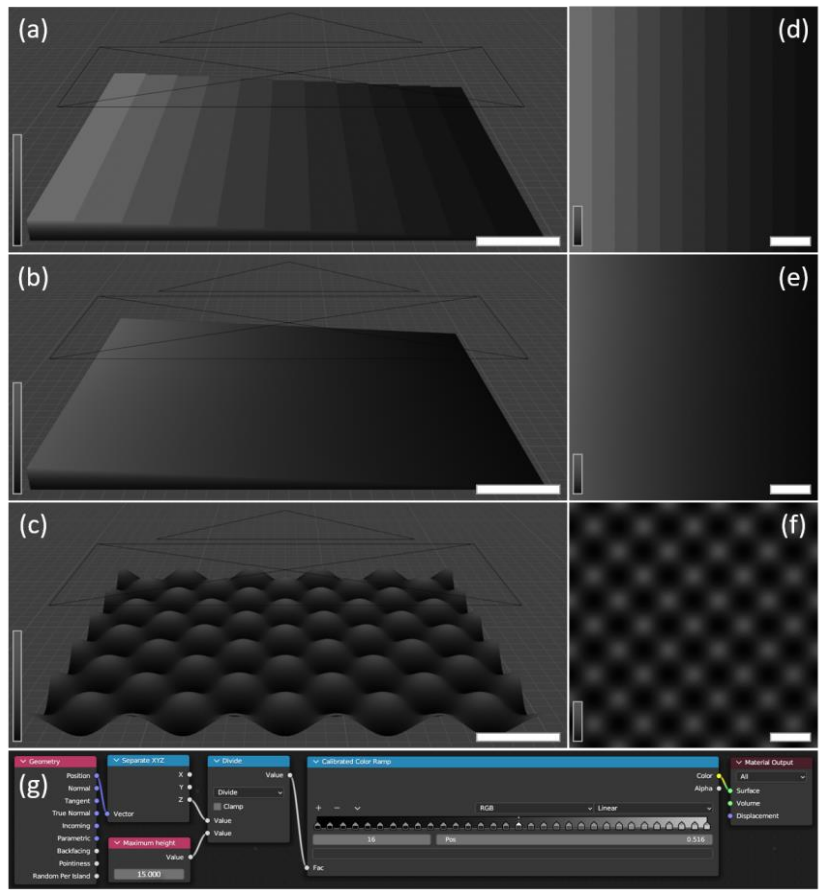

**Figure 3.** Three-dimensional designs of (a) stairs, (b) ramp, and (c) sinusoidal patterns, shaded according to depth (Z-axis). Scale bars represent 40 µm. (d, e, f) Rendered images of each model. Scale bars represent 40 µm. (g) Example of shading node setup for calibrating a maximum depth of 15 µm. Color is assigned based on Z-axis position, with black representing 0 µm and light gray representing 15 µm of material removal.

## 2.4. Characterization

The surface topography of photoresist molds and PMDS replicates was characterized by profilometry, scanning electron microscopy (SEM), and confocal microscopy. The Dektak profiler (Bruker) was used to scan each sample with a tip radius of 2 µm at a speed of 20 µm/s. The obtained profiles were compared with the respective designs based on the difference between the experimental data and the expected depth (residuals). The root mean square (RMS) of the residuals was also calculated, following the analysis from [13] in the general design and in different segments.

The surface of the photoresist molds and PDMS replicates was observed by SEM Quanta 400 (FEI) using its secondary electron detector while the samples were tilted at 30°. Samples were coated with approximately 20 nm of gold via metallization to avoid the accumulation of charge and image artifacts. 3D models of the samples were obtained using the z-stack feature of the LSM 800 (Laser Scanning Microscope, ZEISS) upright configuration. The fluorescent nature of the photoresist allowed for direct imaging without additional sample preparation.

## 3. Results and discussion

In Figure 4, we present the characterization of the exposition of ramp and stairs designs. Figure 4a and 4d show the confocal images of the developed photoresist. In figures 4b and 4e, is shown the profilometer measurements (experimental data) for both ramp and stair designs. For comparison, the expected result (obtained from the inversion of the calibration curve of the respective design) is presented. The difference between these results is represented by residual curves. For all exposures analyzed by profilometry, the final transition between the lowest point and unexposed surface was not considered because of the accuracy of the profilometer tip size. Additionally, figures 4c and 4f present an image of the result obtained by SEM is shown to demonstrate the accordance between the design and the developed overall structure.

The ramp design (Figures 4a, b, and c) demonstrated remarkable linearity in the descent profile achieved in the initial exposure without requiring spatial correction beyond calibration. While depth variations were observed, the residual depth error within any analyzed region remained below 2.5 µm. Although not the primary focus of this study, these depth discrepancies can be readily addressed through recalibration or iterative methods as outlined in [13]. Lateral dimensions accurately matched the design specifications. Overall exposure quality, assessed using RMS, was 1.00 µm for the entire ramp. A more granular analysis revealed RMS values ranging from 0.10 µm at the upper third of the descent to 1.61 µm at the ramp's base.

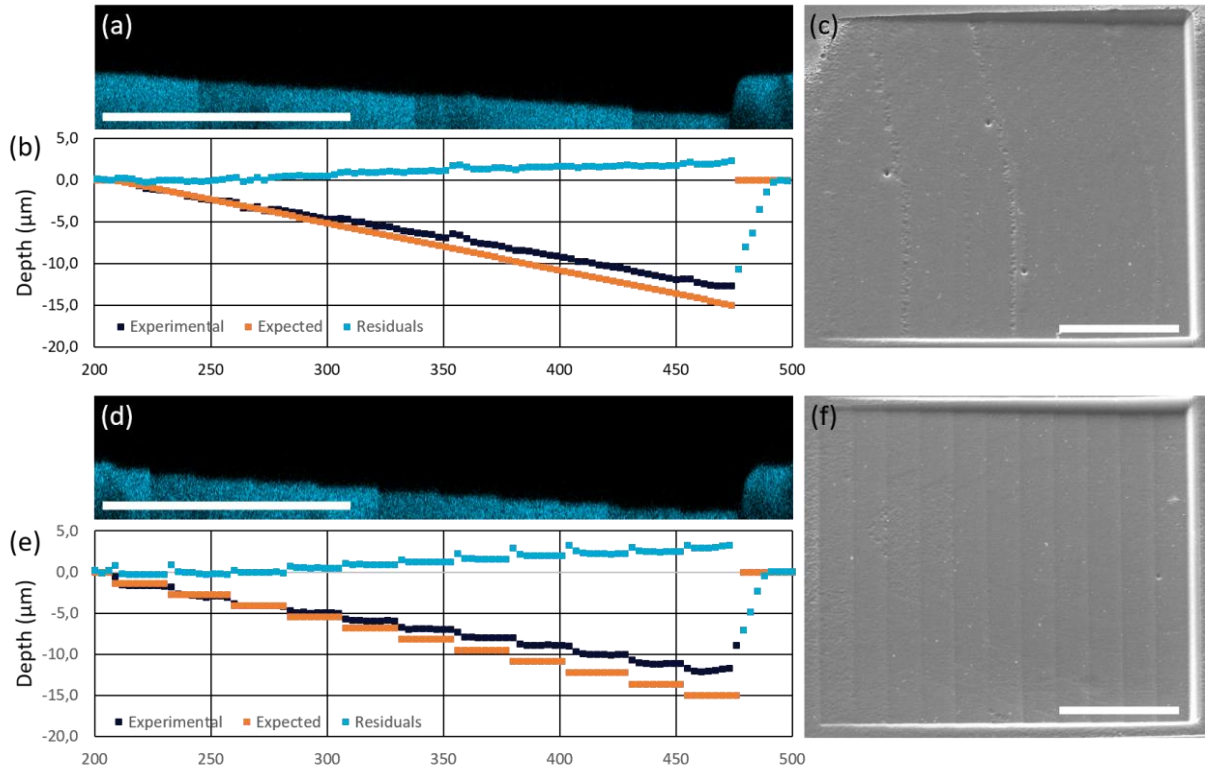

**Figure 4.** (a, d) Confocal images of exposed ramp and stair designs, respectively, based on autofluorescence. (b, e) Comparison of experimental data, expected results, and their corresponding residuals for ramp and stair designs, respectively. (c, f) SEM images of exposed ramp and stair designs tilted at 30°. Scale bars: 100 µm.

The stair design exhibited equivalent results, as depicted in Figures 4d, 4e, and 4f. Consistent stair height and width were observed across the entire exposed area. As anticipated, the stair level edges exhibited rounded shapes due to lateral photon diffusion during exposure. This phenomenon arises from the lateral light beam intensity distribution, causing dose transitions at sharp design edges and subsequent lateral development.

While proximity exposure correction (PEC) could potentially mitigate this issue through iterative adjustments, as discussed in [13], its application is primarily focused on focused beam (electron or laser) exposure rather than large-area exposure. Quantitatively, the RMS value ranged from 3.07 µm at the deepest point to 0.05 µm at the second step level, with an overall average of 1.43 µm.

To further demonstrate the method's capabilities, we fabricated more complex structures: microlenses with a 30 µm diameter and sinusoidal surfaces with varying frequencies. Figure 5 presents the generated digital masks and corresponding images of the produced surfaces. The resulting structures exhibited heights below 15 µm (approximately 7 µm), consistent with previous ramp and stair results. Nevertheless, the 3D design's dimensions and slopes were accurately reproduced. Figures 5a-c display a microlens array characterized by round domes and consistent periodicity. Similarly, Figures 5d-i showcase the regularity and extensive coverage of sinusoidal surfaces with identical heights but doubled frequencies. In both cases, the precise 3D design translated into accurate digital mask instructions, leading to successful fabrication of the target surfaces.

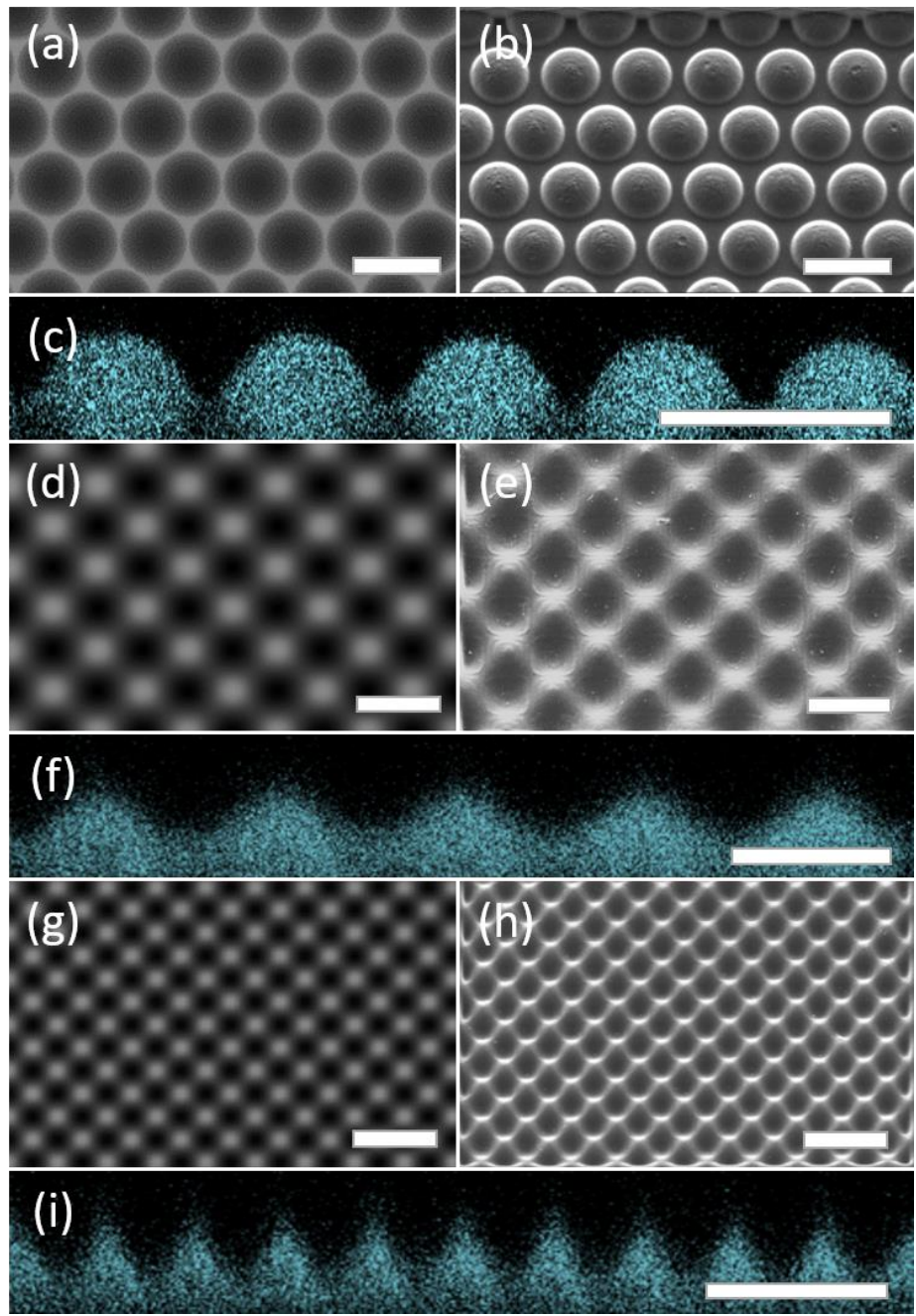

**Figure 5.** (a, d, g) Rendered designs of microlenses, sinusoidal surfaces, and doubled-frequency sinusoidal surfaces, respectively, generated using the proposed method. (b, e, h) Corresponding SEM images of developed surfaces tilted at 30°. (c, f, i) Corresponding confocal microscopy images of developed surfaces in orthogonal view. Scale bars: 50 μm.

A variety of geometric shapes, including pyramids, cones, cylinders, cubes, and hemispheres, were fabricated with dimensions ranging from 5 to 15 μm in side length, diameter, and height. These shapes were chosen to represent a diverse set of design elements, encompassing sharp tips (pyramids, cones), edges (cylinders, cubes), linear slopes (pyramids, cones), and smooth curves (hemispheres). By combining these structures, we aimed to demonstrate the versatility of our approach for a wide range of design possibilities. Figure 6 presents the 3D designs, corresponding digital masks, and resulting microstructures as observed by electron microscopy.

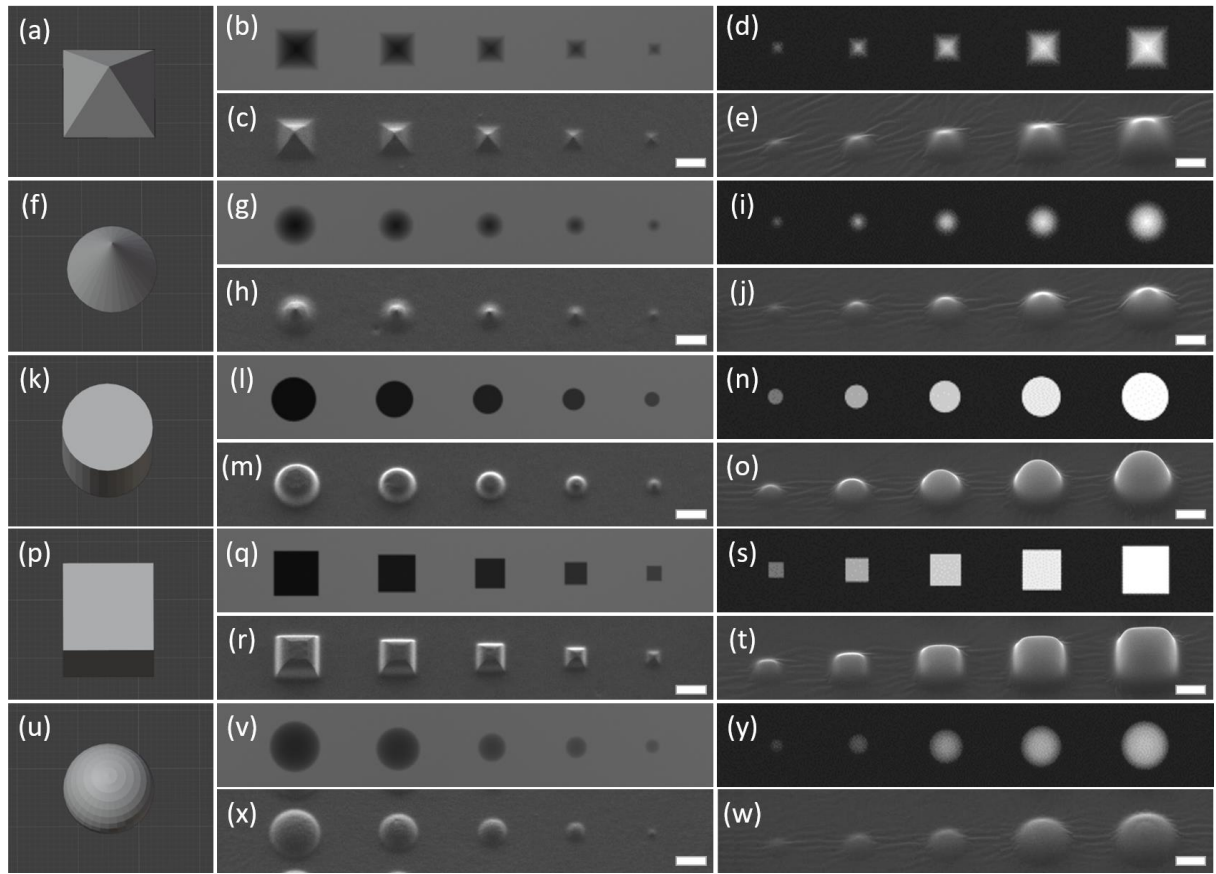

**Figure 6.** (a-e) Pyramid, (f-j) cone, (k-o) cylinder, (p-t) cube, and (u-w) hemisphere. (a, f, k, p, u) 3D designs. (b, g, l, q, v) Generated grayscale masks for direct lithography (5-15 µm). (c, h, m, r, x) SEM images of fabricated structures (20° tilt). (d, i, n, s, y) Generated grayscale masks for PDMS molding (5-15 µm). (e, j, o, t, w) SEM images of PDMS replicates (20° tilt). Scale bars: 10 µm.

A comprehensive analysis of the figures reveals accurate reproduction of all dimensions, with consistent material removal on flat base levels. As previously observed, the calibrated process maintains linear material removal on pyramid and cone sides (Figure 5c, f). The proximity effect, previously noted in stair designs, is evident in cylinders and cubes (Figure 5i, l), resulting in top sections smaller than the base. This effect is more pronounced in taller structures due to increased contrast between dark tops and light bottom, causing photon leakage on walls (Figure 5h, k). While vertical wall fabrication remains a known lithography challenge, the proposed corrections and parameters can address this issue (ref). The final set of figures demonstrates well-defined curved domes on hemispherical shapes (Figure 5o). Given the simultaneous production of all structures, this approach enables the fabrication of any 3D idealized design through the calibrated rendering protocol outlined in this work.

## 4. Conclusions

This work presents a simplified method for generating initial image files for grayscale lithography. We demonstrated the utility of the free, open-source digital rendering software Blender for designing 3D microstructures, assigning color to geometrical heights based on calibrated depth, and producing image files via top-view rendering. For simple structures like stairs and ramps, our method accurately reproduced the overall shape compared to the original design. We extended this approach to complex 3D sinusoidal surfaces, achieving similar results. Given the increasing accessibility and user-friendliness of digital rendering tools, this study offers a promising approach to rapidly generating initial designs with minimal effort.

**Conflict of interest:**

The authors have no conflicts of interest to declare.

**Authors contribution:**

Fabrício Frizera Borghi: Conceptualization; Investigation; Data curation; Formal analysis; Funding acquisition; Investigation; Methodology; Original draft.

Mohammed Bendimerad: Formal analysis; Investigation; Original draft.

Marie-Ly Chapon: Formal analysis; Investigation; Writing - review & editing.

Tatiana Petithory: Formal analysis; Investigation; Writing - review & editing.

Laurent Vonna: Funding acquisition; Project administration; Resources; Supervision; Writing - review & editing.

Laurent Pieuchot: Conceptualization; Funding acquisition; Project administration; Resources; Supervision; Writing - review & editing.

**Acknowledgments**

We acknowledge the people running the IS2M platforms, the Agence Nationale de la Recherche, the Centre National de la Recherche Scientifique and the Ministère de l'enseignement supérieur et de la recherche for financial support. This study was financed in part by the Coordenação de Aperfeiçoamento de Pessoal de Nível Superior - Brasil (CAPES) [Finance Code 001]. This work of the Interdisciplinary Institute HiFunMat, as part of the ITI 2021–2028 program of the University of Strasbourg, CNRS, and Inserm, was supported by IdEx Unistra [ANR-10-IDEX-0002] and SFRI STRAT'US project [ANR-20-SFRI-0012] under the framework of the French Investments for the Future Program.